\documentclass[conference]{IEEEtran}
\IEEEoverridecommandlockouts
\usepackage{cite}
\usepackage{amsmath,amssymb,amsfonts}
\usepackage{algorithmic}
\usepackage{graphicx}
\usepackage{textcomp}
\usepackage{xcolor}
\usepackage{hyperref}
\def\BibTeX{{\rm B\kern-.05em{\sc i\kern-.025em b}\kern-.08em
    T\kern-.1667em\lower.7ex\hbox{E}\kern-.125emX}}
\begin{document}

\title{MediHunt: A Network Forensics Framework for Medical IoT Devices\\
}

\author{
\IEEEauthorblockN{Ayushi Mishra}
\IEEEauthorblockA{\textit{Computer Science and Engineering} \\
\textit{Indian Institute of Technology}\\
Kanpur, India \\
ayushim@cse.iitk.ac.in}
\and
\IEEEauthorblockN{ Tej Kiran Boppana}
\IEEEauthorblockA{\textit{Computer Science and Engineering} \\
\textit{Indian Institute of Technology}\\
Kanpur, India \\
tejkiranb@cse.iitk.ac.in}
\and
\IEEEauthorblockN{Priyanka Bagade}
\IEEEauthorblockA{\textit{Computer Science and Engineering} \\
\textit{Indian Institute of Technology}\\
Kanpur, India \\
pbagade@cse.iitk.ac.in}
}

\maketitle

\begin{abstract}
The Medical Internet of Things (MIoT) has enabled small, ubiquitous medical devices to communicate with each other to facilitate interconnected healthcare delivery. These devices interact using communication protocols like MQTT, Bluetooth, and Wi-Fi. However, as MIoT devices proliferate, these networked devices are vulnerable to cyber-attacks. This paper focuses on the vulnerabilities present in the Message Queuing Telemetry and Transport (MQTT) protocol. The MQTT protocol is prone to cyber-attacks that can harm the system's functionality. The memory-constrained MIoT devices enforce a limitation on storing all data logs that are required for comprehensive network forensics. This paper solves the data log availability challenge by detecting the attack in real-time and storing the corresponding logs for further analysis with the proposed network forensics framework: MediHunt. Machine learning (ML) techniques are the most real safeguard against cyber-attacks. However, these models require a specific dataset that covers diverse attacks on the MQTT-based IoT system for training. The currently available datasets do not encompass a variety of applications and TCP layer attacks. To address this issue, we leveraged the usage of a flow-based dataset containing flow data for TCP/IP layer and application layer attacks. Six different ML models are trained with the generated dataset to evaluate the effectiveness of the MediHunt framework in detecting real-time attacks. F1 scores and detection accuracy exceeded 0.99 for the proposed MediHunt framework with our custom dataset.

\end{abstract}


\section{Introduction}
\begin{figure}[ht]
\centering 
\includegraphics[width=0.5\textwidth]{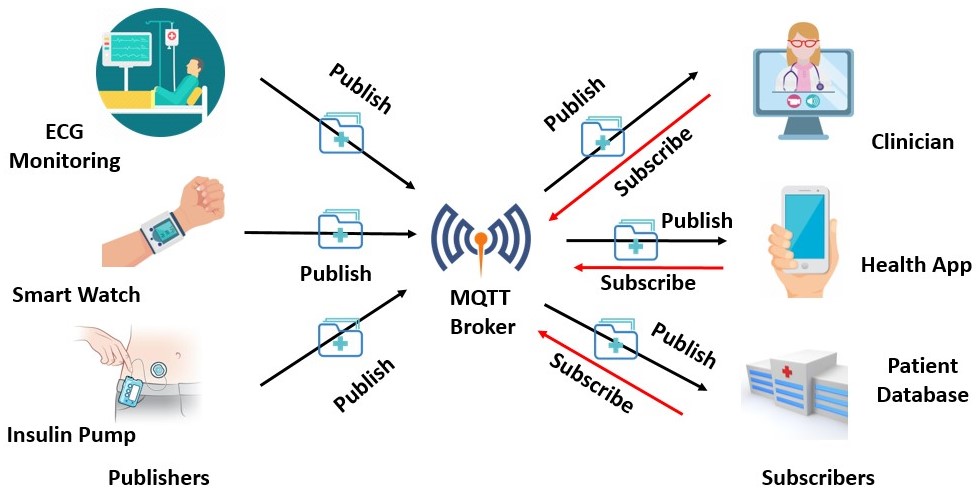}
\caption{MQTT-based Medical IoT System}
\label{fig:mqtt}
\end{figure}

Integrating the Medical Internet of Things (MIoT)\cite{fl} into the healthcare industry enables effective patient care by connecting medical devices, wearables, and clinicians for various healthcare applications. These MIoT systems utilize several communication protocols like Wi-Fi, Bluetooth, MQTT, and Zigbee to enable efficient and reliable data exchange. As an example, in this paper, we have considered a smart hospital setup as shown in figure~\ref{fig:mqtt}. It uses the lightweight Message Queuing Telemetry Protocol (MQTT) to establish communication between medical IoT devices and wearables acting as publishers and central systems acting as subscribers, including the clinician computer, patient health app, and medical database system. The publisher sends the patient data as a topic to the MQTT broker. The broker routes these topics to the subscribers who subscribe to these topics to receive the patient data of interest.

\begin{figure*}[ht] 
\centering 
\includegraphics[width=13cm,height=8cm]{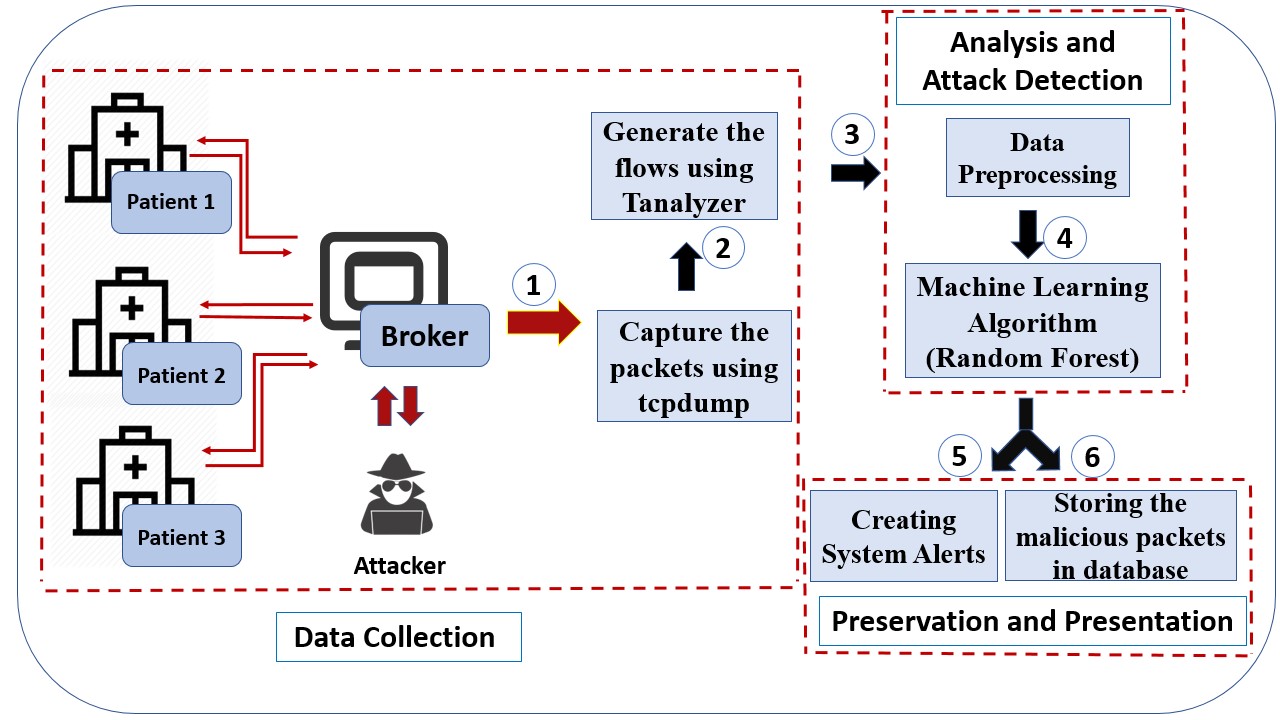}
\caption{MediHunt: Framework for MQTT Network Forensics}
\label{fig:framework}
\end{figure*}

The MQTT protocol in a smart hospital setting is suitable for transmitting data over low-bandwidth networks and low-power or intermittent connections. It relays real-time patient vital signs, monitor equipment status, track inventory, and facilitate communication between different medical devices with low overhead. However, MIoT's tight integration of software and hardware makes it possible for cyber-attacks to have catastrophic physical effects. These MIoT systems are typically attacked using data theft, traffic assaults, and denial of service attacks\cite{privacy1}. Therefore, it is essential to analyze the network traffic to detect malicious activities and generate insights to find the affected areas of the system. 

Our work presents MediHunt, an automated MQTT-based network forensics framework that helps to analyze the network traffic with suspicious activities. It detects intrusions and stores malicious traffic in the database for forensic analysis. In our previous work \cite{tej}, we have created a flow-based network traffic dataset for the prominent MQTT network attacks such as TCP/IP, brute-force, malformed packets, and port scanning attacks. The state-of-the-art network attack detection methods \cite{mqtt_dos_attack}\cite{liu2021bayesian}\cite{ciklabakkal2019artemis} use packet-based flows to simulate the network traffic. The packet-based method only provides the data communicated between various devices, which is insufficient to perform the forensics on the system to trace the malicious activities. The flow-based methods contain the host IP address, number of packets sent, network protocol, and port details. This extensive information can help strengthen the forensics analysis to detect the affected areas of the system as well as avoid such attacks on mission-critical MIoT devices. The proposed MediHunt framework is evaluated by training six different ML models, namely Decision Tree (DF), Random Forest (RF), Support Vector Machine (SVM), Naive Bayes (NB), Multilayer Perceptron (MLP), and XGBoost (XGB), on the generated dataset for real-time attack detection. 

The rest of the paper is organized as follows: Section 2 provides an overview of the related works on attack detection in MQTT networks. Section 3 introduces the proposed MediHunt network forensic framework for MQTT networks. Section 4 explains the experimental setup used to generate the MQTT attack dataset. Section 5 explains the ML model's training and its results. Section 6 discusses the evaluation of the MediHunt framework on Raspberry Pi. Finally, we concluded our work in Section 7.

\section{Related Work}

The network forensics system can generate evidence from the affected areas of the MIoT systems. The network intrusion detection system(NIDS) is an integrated part of network forensics that detects malicious traffic and behavior. Mishra et al.\cite{nf_iot} proposed a network forensics framework to detect DoS attacks on Wi-Fi-configured Raspberry Pi. The evidence was gathered using the network analyzer tool Wireshark which provides the key reinitiation between the Raspberry Pi and Wi-Fi access point. Given the interpretability of rule-based learning, Giuseppe Potrino et al.\cite{potrino2019modeling}, Qi Liu et al.\cite{liu2021bayesian}, Umberto Morelli et al.\cite{mqtt_dos_attack} proposed rule-based learning security systems for mitigating denial of service (DoS) attacks in MQTT networks. The in-depth domain knowledge required by the rule-based techniques makes them challenging to employ in real applications. Also, their work is not focused on detecting active attacks. Ege Ciklabakkal et al.\cite{ciklabakkal2019artemis} treated MQTT attack detection as anomaly detection and used ML models like autoencoder (AE) and one-class SVM (OCSVM) to detect attack packets as anomalies. 

Hector Alaiz-Moreton et al.\cite{multiclass_detection} in their work also generated a public MQTT-specific dataset and trained XGBoost, deep recurrent models LSTM, and GRU on their dataset. Though their dataset contains diverse attacks, recurrent models are generally too slow for real-time traffic analysis. But, \cite{ciklabakkal2019artemis}, \cite{multiclass_detection} have used packet-based features to train ML models on their dataset. Some packet features extracted, such as MQTT message payload, message length, and topic length, cannot be extracted if the packet payload is encrypted, and inspecting packet payload information might raise privacy concerns. For MIoT devices, Mishra et al.\cite{df_miot} present an intrusion detection system and enforce physiological data modeling. 

The above-mentioned ML-based intrusion detection methods use packet-based data that does not contain the required information about the network traffic to do thorough network forensics. The flow-based data provides network details along with the data exchanged between devices. In the literature review, it is observed that there is an unavailability of publicly available flow-based MQTT-specific datasets to train attack detection systems. The proposed MediHunt is an automatic network forensics framework for real-time detection of MQTT network flow-based traffic attacks. It utilizes flow-based traffic monitoring to detect a variety of TCP/IP layer and application layer attacks on MQTT networks. 

\section{Proposed Network Forensics Framework for MQTT Networks}

This section presents a real-time ML-based network forensic framework, MediHunt to monitor MQTT networks in the smart hospital environment as shown in Figure~\ref{fig:framework}. The MediHunt forensic framework includes data collection, analysis, attack detection, presentation, and preservation of evidence. The data collection involves collecting network traffic data at fixed intervals in packets using tcpdump\cite{tcpdump}. These packets are then converted into flows using Tranalyzer\cite{tranalyzer:online}. The collected flows are then preprocessed to use the ML model for attack detection in the analysis and attack detection phases. To generate network traffic, we created diverse attacks. We have used a Linux machine as the attack launch system on an MQTT network. The network was hampered using six categories of attack: invalid subscribe/publish, TCP/SYN flood, brute force, malformed packet, port scanning, and WILL payload.

\section{ML Model Training for Attack Detection}
Intrusion detection is an integral part of the proposed MediHunt network forensics framework. This section focuses on MQTT network traffic data collection, ML model training, and performance analysis.
\subsection{MQTT Network Traffic Data Collection}
We used the flow-based MQTT dataset\cite{tej} from our previous work to train the intrusion detection ML models. The experimental setup used for flow-based data collection deployed the MQTT network with ten consumers and a broker for exchanging messages. Each client and broker operated on separate virtual machines. The network was operated on the Google Cloud Platform (GCP), where each client is created using PAHO MQTT\cite{pahomqtt} package, and the broker is created using Mosquitto\cite{mosquitto} MQTT broker. Out of the ten consumers, five clients were designated as publishers and five as subscribers. We conducted various attacks, including application layer (MQTT) attacks and TCP/IP layer attacks, to generate a robust training dataset for the network forensic detection system. The packet capture(PCAP) collected was converted to flow-based to obtain the required training data. The number of samples for Benign and Malicious were 55\% and 45\%, respectively. For the attack-wise distribution of flow-based samples, we analyzed a class imbalance problem where the normal flows were quite larger than the diverse attack flows. We used under-sampling and oversampling techniques to address this issue.

\begin{figure} 
\centering
\includegraphics[scale=0.82]{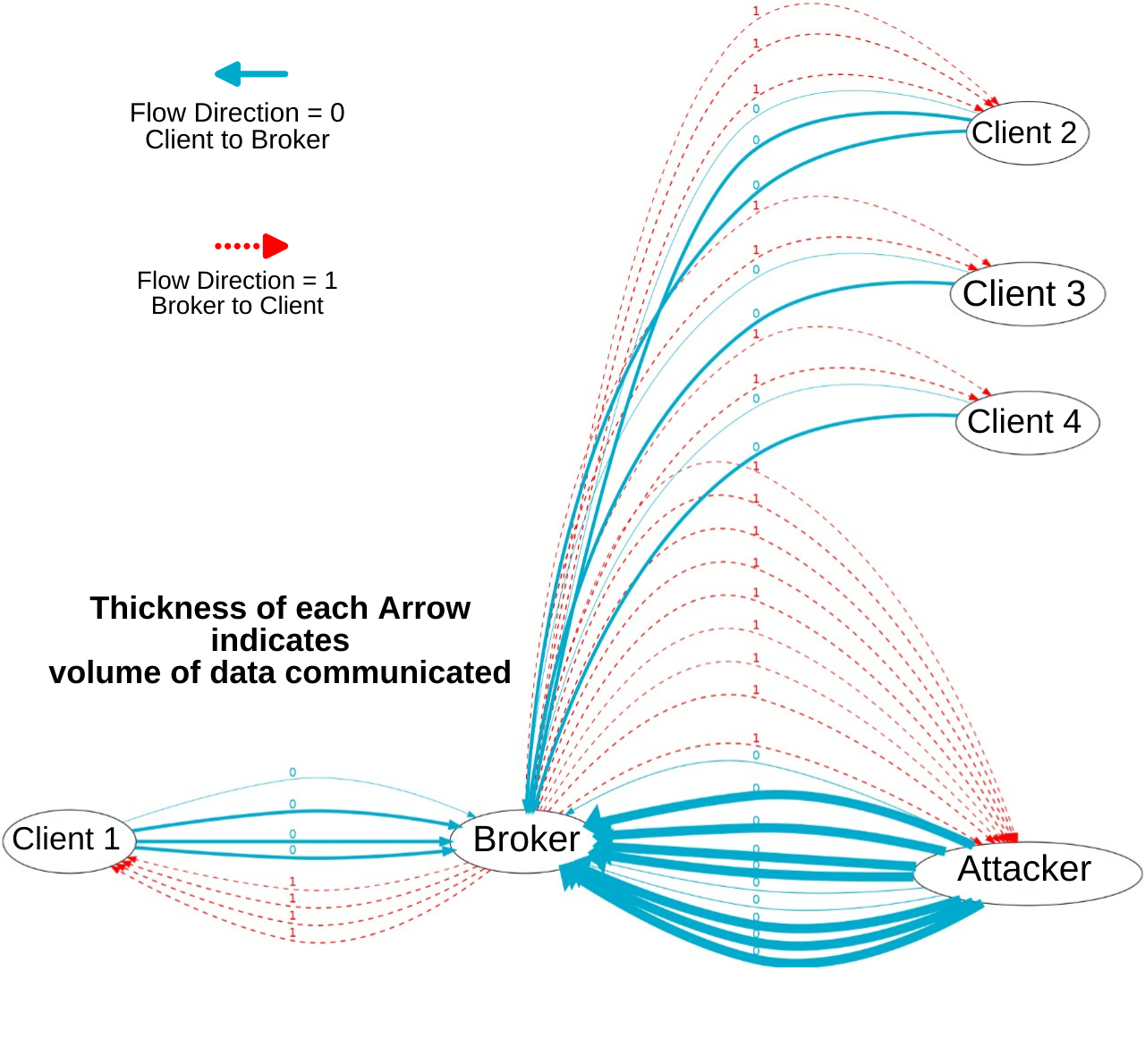}
\caption{MQTT WILL Payload Attack}
\label{fig:will}
\end{figure}

The state of the MQTT network under WILL Payload attacks is shown in Figure~\ref{fig:will}. The figure illustrates traffic between various nodes in the network during attacks. Each arrow in the figure represents a flow of packets from source to destination. The solid arrows represent the flow direction from the client to the broker, and the flow direction is designated by value zero. In contrast, the dotted arrows represent the flow direction from the broker to the client, and the flow direction is designated by value one. The thickness of the arrows indicates the volume of data (in bytes) that was communicated.



\label{sub:data_preprocessing}

\subsection{ML Model Training and Performance Analysis}

\begin{figure} 
\centering
\includegraphics[scale=0.15]{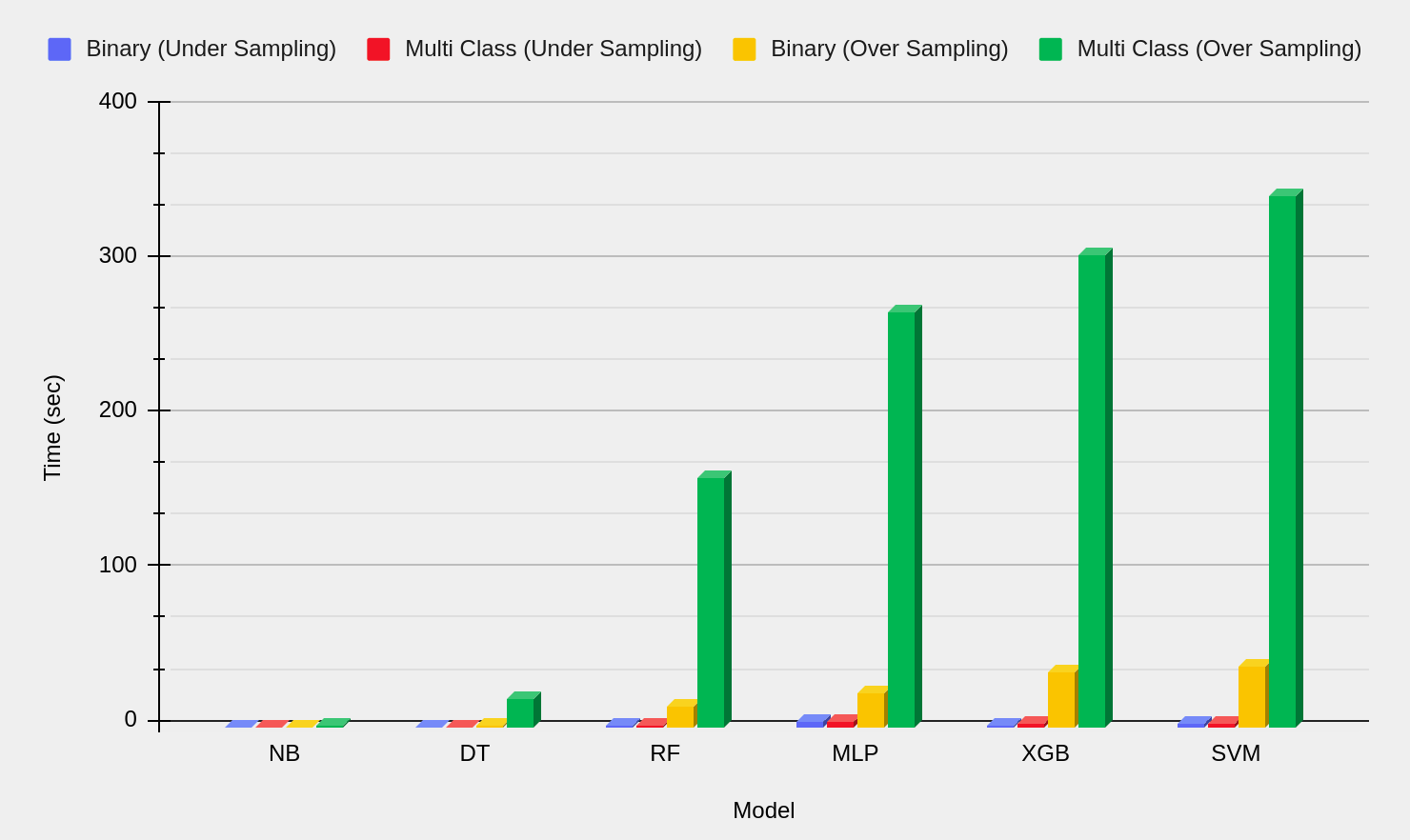}
\caption{Training Times of ML models}
\label{fig:time}
\end{figure}

We have trained six different ML classifiers, namely Decision Tree (DT), Random Forest (RF), Support Vector Machine (SVM), Gaussian Naive Bayes (NB), Multi-Layer Perceptron (MLP), and XGBoost (XGB). We have trained the ML models to perform binary and multi-class classification on the under-sampled and over-sampled training data. In binary classification, a flow is classified as benign/malicious, and in multi-class classification. The classification results on the test data are shown in Table \ref{table4}. Random Forest algorithms outperform binary and multi-class classification with an F1 score of 1. The training time and classification results are comparable for both under-sampled and over-sampled datasets in the case of binary classification. Figure~\ref{fig:time} illustrates that Naive Bayes takes the least training time among all the other ML models. 



\begin{table}
\centering
\caption{Classification Result}
\label{table4}
\begin{tabular}{|p{1cm}|p{1.25cm}|p{1.2cm}|p{1cm}|p{1.3cm}|}
\hline
\multicolumn{5}{|c|}{\textbf{Multi Class Classification (Under Sampling)}}\\
\hline
\textbf{Model}&\textbf{Accuracy} &\textbf{Precision} &\textbf{Recall}&\textbf{F1-Score}\\
\hline
DT & 0.9081& 0.9071& 0.9081& 0.9075\\
\hline
RF &0.9323& 0.9414& 0.9323& \textbf{0.9349}\\
\hline
SVM &0.8531&0.8438&0.8531&0.8154\\
\hline
NB &0.8042&0.8602& 0.8042& 0.7957\\
\hline
MLP &0.8848& 0.9247& 0.8848& 0.8813\\
\hline
XGB & 0.9261& 0.9366& 0.9261& \textbf{0.9293}\\
\hline\hline
\multicolumn{5}{|c|}{\textbf{Multi Class Classification (Over Sampling)}}\\
\hline
\textbf{Model}&\textbf{Accuracy}&\textbf{Precision}&\textbf{Recall}&\textbf{F1-Score}\\
\hline
DT & 0.9954& 0.9953& 0.9954& 0.9953\\
\hline
RF &0.9962& 0.9970& 0.9962& \textbf{0.9963}\\
\hline
SVM & 0.9891& 0.9920& 0.9891& 0.9905\\
\hline
NB &0.9853& 0.9876& 0.9853& 0.9848\\
\hline
MLP &0.9955& 0.9970& 0.9955& 0.9956\\
\hline
XGB & 0.9968& 0.9972& 0.9968& \textbf{0.9969}\\
\hline\hline
\multicolumn{5}{|c|}{\textbf{Binary Classification (Under Sampling)}}\\
\hline
\textbf{Model}&\textbf{Accuracy}&\textbf{Precision}&\textbf{Recall}&\textbf{F1-Score}\\
\hline
DT & 0.9993&0.9993&0.9993&0.9993\\
\hline
RF & 0.9995& 0.9995& 0.9995& \textbf{0.9995}\\
\hline
SVM & 0.9973& 0.9973& 0.9973& 0.9973\\
\hline
NB &0.9840& 0.9841& 0.9840& 0.9841\\
\hline
MLP & 0.9990& 0.9990& 0.9990& 0.9990\\
\hline
XGB & 0.9994& 0.9994& 0.9994& \textbf{0.9994}\\
\hline\hline
\multicolumn{5}{|c|}{\textbf{Binary Classification (Over Sampling)}}\\
\hline
\textbf{Model}&\textbf{Accuracy}&\textbf{Precision}&\textbf{Recall}&\textbf{F1-Score}\\
\hline
DT & 1.0&1.0&1.0&1.0\\
\hline
RF &1.0&1.0&1.0&\textbf{1.0}\\
\hline
SVM & 0.9993& 0.9993& 0.9993& 0.9993\\
\hline
NB &0.9981&0.9981& 0.9981& 0.9981\\
\hline
MLP &0.9999&0.9999&0.9999&0.9999\\
\hline
XGB & 1.0 & 1.0 & 1.0 & \textbf{1.0}\\
\hline
\end{tabular}
\end{table}

\section{Evaluation on Raspberry Pi}

\begin{figure} 
\centering
\includegraphics[scale=0.10]{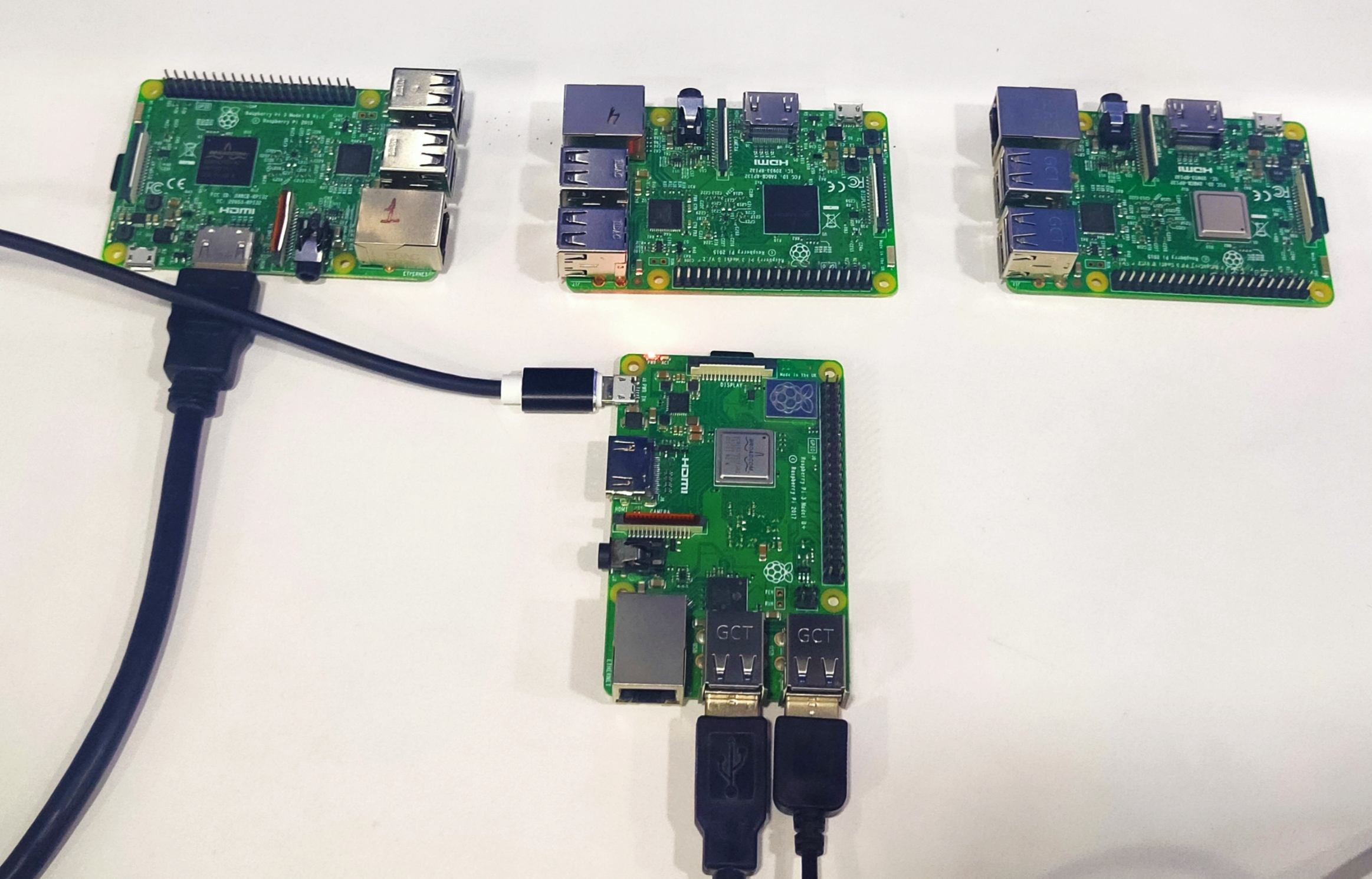}
\caption{Network Forensics Framework Deployed on Raspberry Pi Devices}
\label{fig:rasp}
\end{figure}

To implement our network forensics framework on MIoT devices with limited resource availability, we analyzed ML algorithms on a Raspberry Pi. As depicted in figure~\ref{fig:rasp}, the models were trained and evaluated using four Raspberry Pi 3B models. In addition, we determined that the inference and training times of the ML algorithms were comparable. The inference time on the cloud platform was around 2ms, which in Raspberry Pi was 0.17ms. The proposed MediHunt network forensics framework is a lightweight intrusion detection system solution that can be readily deployed on any resource-constrained and low-computing MIoT device.


\section{Conclusion}

This paper proposes designing and implementing a real-time network forensic framework MediHunt for MIoT MQTT networks. MediHunt integrates ML algorithms to detect possible attacks on the MQTT brokers. It uses custom-built flow-based MQTT network traffic data containing diverse attacks. The currently 
available packet-based intrusion detection systems are unsuitable for building real-time attack detection systems due to the higher latency in dissecting each packet. To address this issue, we used a flow feature-based dataset containing diverse attacks possible on MQTT networks from our previous work. Flow-based features are generated by summarizing statistical and header information from related packets. Since flow data is summarized, fewer resources are needed to analyze it. A comparative analysis of the performances of various machine learning algorithms is provided and achieved a detection accuracy of 0.99. We evaluated our framework on Raspberry Pi devices to show the feasibility of its working in resource-constraint MIoT devices.

\bibliographystyle{IEEEtran}
\bibliography{mybibliography.bib}

\begin{thebibliography}{10}
\providecommand{\url}[1]{#1}
\csname url@samestyle\endcsname
\providecommand{\newblock}{\relax}
\providecommand{\bibinfo}[2]{#2}
\providecommand{\BIBentrySTDinterwordspacing}{\spaceskip=0pt\relax}
\providecommand{\BIBentryALTinterwordstretchfactor}{4}
\providecommand{\BIBentryALTinterwordspacing}{\spaceskip=\fontdimen2\font plus
\BIBentryALTinterwordstretchfactor\fontdimen3\font minus \fontdimen4\font\relax}
\providecommand{\BIBforeignlanguage}[2]{{%
\expandafter\ifx\csname l@#1\endcsname\relax
\typeout{** WARNING: IEEEtran.bst: No hyphenation pattern has been}%
\typeout{** loaded for the language `#1'. Using the pattern for}%
\typeout{** the default language instead.}%
\else
\language=\csname l@#1\endcsname
\fi
#2}}
\providecommand{\BIBdecl}{\relax}
\BIBdecl

\bibitem{fl}
A.~Mishra, S.~Saha, S.~Mishra, and P.~Bagade, ``A federated learning approach for smart healthcare systems,'' \emph{CSI Transactions on ICT}, pp. 1--6, 2023.

\bibitem{privacy1}
\BIBentryALTinterwordspacing
A.~Nieto, R.~Rios, and J.~Lopez, ``Iot-forensics meets privacy: Towards cooperative digital investigations,'' \emph{Sensors}, vol.~18, no.~2, 2018. [Online]. Available: \url{https://www.mdpi.com/1424-8220/18/2/492}
\BIBentrySTDinterwordspacing

\bibitem{tej}
\BIBentryALTinterwordspacing
T.~K. Boppana and P.~Bagade, ``Gan-ae: An unsupervised intrusion detection system for mqtt networks,'' \emph{Engineering Applications of Artificial Intelligence}, vol. 119, p. 105805, 2023. [Online]. Available: \url{https://www.sciencedirect.com/science/article/pii/S0952197622007953}
\BIBentrySTDinterwordspacing

\bibitem{mqtt_dos_attack}
\BIBentryALTinterwordspacing
U.~Morelli, I.~Vaccari, S.~Ranise, and E.~Cambiaso, ``Dos attacks in available mqtt implementations: Investigating the impact on brokers and devices, and supported anti-dos protections,'' in \emph{The 16th International Conference on Availability, Reliability and Security}, ser. ARES 2021.\hskip 1em plus 0.5em minus 0.4em\relax New York, NY, USA: Association for Computing Machinery, 2021. [Online]. Available: \url{https://doi.org/10.1145/3465481.3470049}
\BIBentrySTDinterwordspacing

\bibitem{liu2021bayesian}
Q.~Liu, H.~B. Keller, and V.~Hagenmeyer, ``A bayesian rule learning based intrusion detection system for the mqtt communication protocol,'' in \emph{The 16th International Conference on Availability, Reliability and Security}, 2021, pp. 1--10.

\bibitem{ciklabakkal2019artemis}
E.~Ciklabakkal, A.~Donmez, M.~Erdemir, E.~Suren, M.~K. Yilmaz, and P.~Angin, ``Artemis: An intrusion detection system for mqtt attacks in internet of things,'' in \emph{2019 38th Symposium on Reliable Distributed Systems (SRDS)}.\hskip 1em plus 0.5em minus 0.4em\relax IEEE, 2019, pp. 369--3692.

\bibitem{nf_iot}
A.~Mishra and P.~Bagade, ``Investigating iot systems security attacks using network forensics,'' in \emph{2023 15th International Conference on COMmunication Systems \& NETworkS (COMSNETS)}.\hskip 1em plus 0.5em minus 0.4em\relax IEEE, 2023, pp. 72--77.

\bibitem{potrino2019modeling}
G.~Potrino, F.~De~Rango, and A.~F. Santamaria, ``Modeling and evaluation of a new iot security system for mitigating dos attacks to the mqtt broker,'' in \emph{2019 IEEE Wireless Communications and Networking Conference (WCNC)}.\hskip 1em plus 0.5em minus 0.4em\relax IEEE, 2019, pp. 1--6.

\bibitem{multiclass_detection}
H.~Alaiz~Moreton, J.~Aveleira, J.~Ondicol-Garcia, A.~Muñoz-Castañeda, I.~García, and C.~Benavides, ``Multiclass classification procedure for detecting attacks on mqtt-iot protocol,'' \emph{Complexity}, vol. 2019, pp. 1--11, 04 2019.

\bibitem{df_miot}
A.~Mishra and P.~Bagade, ``Digital forensics for medical internet of things,'' in \emph{2022 IEEE Globecom Workshops (GC Wkshps)}, 2022, pp. 1074--1079.

\bibitem{tcpdump}
P.~Goyal and A.~Goyal, ``Comparative study of two most popular packet sniffing tools-tcpdump and wireshark,'' in \emph{2017 9th International Conference on Computational Intelligence and Communication Networks (CICN)}, 2017, pp. 77--81.

\bibitem{tranalyzer:online}
\BIBentryALTinterwordspacing
tranalyzer. (2022) Tranalyzer. [Online]. Available: \url{https://tranalyzer.com/}
\BIBentrySTDinterwordspacing

\bibitem{pahomqtt}
\BIBentryALTinterwordspacing
mqtt. (2022) Paho-mqtt. [Online]. Available: \url{https://pypi.org/project/paho-mqtt/}
\BIBentrySTDinterwordspacing

\bibitem{mosquitto}
\BIBentryALTinterwordspacing
mosquitto. (2022) Eclipse mosquitto. [Online]. Available: \url{https://mosquitto.org/}
\BIBentrySTDinterwordspacing

\end{thebibliography}

\end{document}